# Environmental Interaction: Recognizing Capability within Everyday Actions

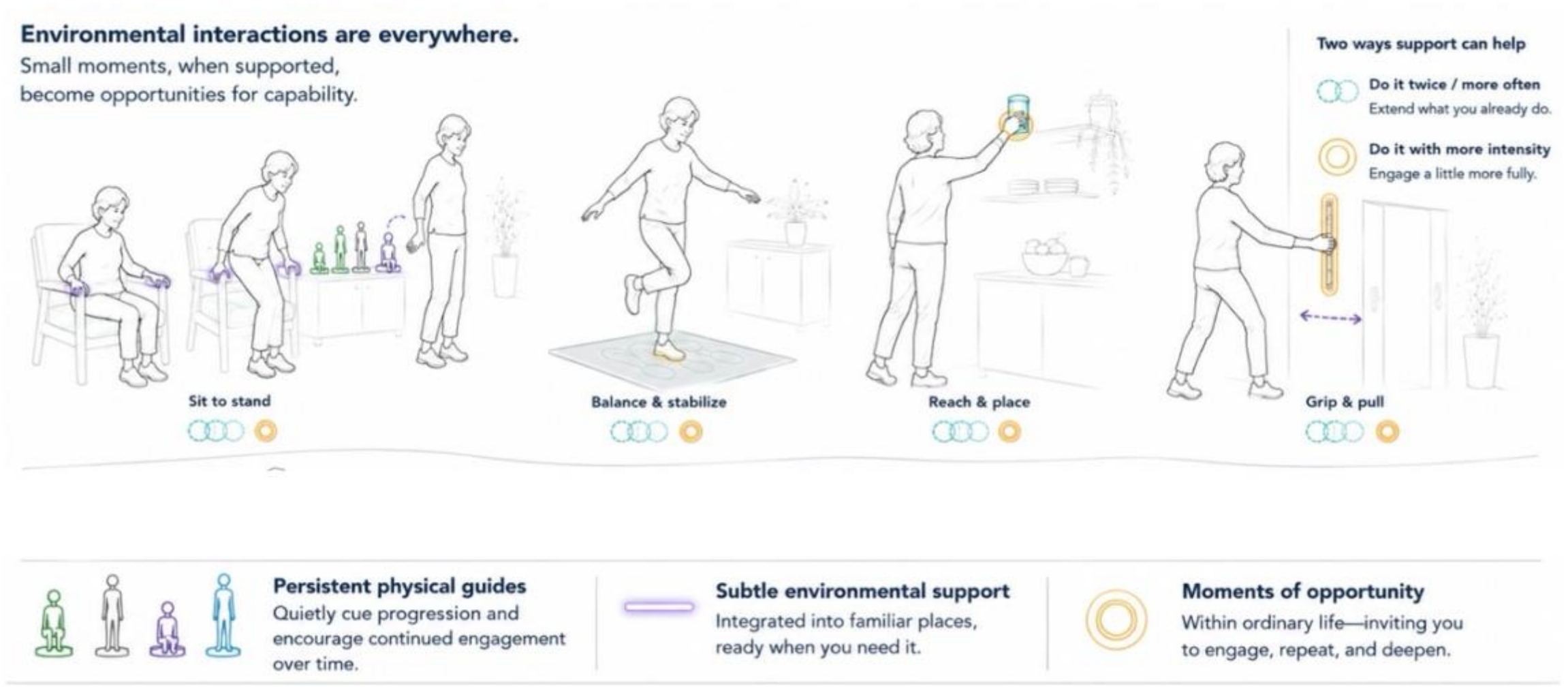

Figure 1: Environmental interaction in practice. The system detects an everyday action already underway — in this case, a sit-to-stand movement — and makes its capability-bearing significance visible, inviting the person to engage further: to repeat the movement, vary its intensity, or simply notice what they are already doing.

M.C. SCHRAEFEL[1], ALEXANDER NG[1], CHRIS TACCA[1], RICHARD GOMER[2], ARTURO VASQUEZ GALVEZ[3], CHRIS FREEMAN[3], ISOBEL THOMPSON[4], MARTIN WARNER[4], JASON ALEXANDER[5]

University of Souththampton UK : 1Wellthlab, 2. DAIS 3. Digital Health, Electronics and Comptuter Science 4 Allied Professionals, School of Health Sciences. University of Bath, UK: 5. Department of Computer Science

**ABSTRACT** Systems designed to support everyday capability consistently ask people to do something additional — track activity, follow programmes. For those who need such support most, this ask is often too high. We propose a different approach: environmental interaction. Through four probe deployments with older adults we show that everyday actions are themselves generative sites for interaction design, and that engagement depends less on sensing and guidance than on revealing in that action that it is polymorphic. A chair sit-stand is also a box squat; a door pull also a row. Each time participants encountered this second frame, the response was surprise, delight, and "I can do that!"

The paper contributes environmental interaction as a design space, a serial probe methodology, and empirically grounded design considerations — including invitation over instruction, recognition before extension, and interaction as presence — centered on recognition as a generative design challenge.

## 1 INTRODUCTION

Interactive systems designed to support movement and health are often experienced as introducing new tasks or additional activities into everyday life. Even when they build on actions people already perform, they typically frame those actions as things to be tracked, increased, or managed, rather than as meaningful sites of interaction in their own right. This raises a question: how might interaction support movement without adding new tasks or interrupting activity *already* underway?

This question is especially important in environments where sedentarism is shaped not only by individual behavior, but by contexts that normalize sedentarism and make movement less likely [7, 84]. Maintaining strength through everyday movement is strongly associated with independence and reduced risk of falls and morbidity in older adults [10, 27]. Yet systems designed to support movement in HCI—from activity trackers to exercise guidance platforms—often focus on monitoring [12, 74], prompting [33, 72], or guiding behavior [39, 63] through dedicated programmes or interfaces. Even where these systems build on existing activity, they rarely support the reinterpretation of everyday actions themselves as opportunities for strength and capability in situ.

This gap became concrete in our co-design work with older adults (Section 3), who expressed a strong desire to maintain strength and independence, yet found structured exercise programmes difficult (or uninteresting) to sustain. What participants described wanting was not new exercises, but support for recognizing and extending the value of actions already embedded in their routines. Simple reframings such as standing up from a chair twice rather than once were immediately intelligible: no new equipment, no new routine, just a small extension of what was already happening. Our insight was that strength practice might be supported not by adding new behaviors, but by reinterpreting existing ones as opportunities for strength practice in their own right — though as the probes would reveal, establishing that reframing for participants themselves proved the deeper and less anticipated challenge.

We term this approach environmental interaction: a design space in which actions already underway in everyday environments are treated as sites of interaction rather than as behaviors to be tracked or replaced. This shifts the locus of interaction from designed artefacts, monitoring systems, or dedicated interfaces, to the actions people are already performing. Where tangible and embodied interaction center engagement around designed artefacts, and monitoring systems sense and act upon behavior, environmental interaction focuses on how existing action becomes recognizable and interactionally meaningful in context. In this framing, interaction occurs when an action already underway becomes openly available — when standing from a chair is no longer only transit between positions, but simultaneously a strength move, a choice, a moment that can be extended or simply noticed and passed. A chair sit-stand is also a box squat; pulling a door open is also a row. These actions are already happening. In this framing, the role of the system is not to add a new task, but to make the capability-bearing potential of existing action perceptible (Figure 1). To explore this idea in practice, we developed sensing-enabled materials embedded within ordinary domestic contexts, allowing actions already underway to be recognized and, where participants chose, extended in situ. We introduce environmental interaction as a design space — one instantiated here through movement and strength-building, but defined independently of that domain. The design challenge it names applies wherever everyday actions carry capability-bearing potential - — the significance an action already carries beyond its immediate function - that goes unrecognized

Any exploration of environmental interaction raises a methodological challenge: because this form of interaction unfolds within ongoing activity and environments that carry their own meanings and routines, conventional evaluation focused solely on predefined outcomes is insufficient to capture how it is understood and taken up in practice. Any exploration of environmental interaction raises a methodological challenge: because this form of interaction unfolds within ongoing activity and environments that carry their own meanings and routines, conventional evaluation focused solely on predefined outcomes is insufficient to capture how it is understood and taken up in practice. We therefore adopted a serial probe approach, using a sequence of in-situ deployments to investigate this space — an approach elaborated in the contributions below."

The contribution of this paper is fourfold:

1. We introduce environmental interaction as a design space, instantiated here to support movement and strength-building within everyday environments.
2. We demonstrate how a sequence of probe-based deployments can function as a serial inquiry into this space, revealing how everyday actions are reinterpreted, taken up, and adapted within the routines of the home.
3. We show empirically that recognition of the capability-bearing significance of everyday action cannot be assumed, identifying recognition as a distinct design challenge for environmental interaction. Across four probes, participants repeatedly expressed surprise when familiar actions were reframed as meaningful capability-building opportunities, suggesting that recognition itself requires active support.
4. We identify a set of empirically grounded design considerations for environmental interaction, including invitation rather than instruction, recognition before extension, interaction as presence rather than event, separation of doing and reflecting, and expanding the option space over time.

This paper presents this work in the following sections: Section 2 reviews related work across movement support, interactive environments, and probe methodologies. Section 3 presents the design rationale and Section 4 the four probe deployments. Section 5 synthesizes findings into design considerations, Section 6 discusses limitations, Section 7 future directions, and Section 8 concludes.

## 2 RELATED WORK

Environmental interaction sits at the intersection of several active areas in HCI, each of which frames interaction differently, and each of which leaves a related space largely unexamined. Together, these strands establish both the conceptual framing of environmental interaction and the basis for its empirical investigation.

### 2.1 Supporting movement in HCI

Human–Computer Interaction has long engaged with the challenge of supporting movement, particularly in domains relating to health, wellbeing, and ageing. A substantial body of work has focused on designing systems and interfaces to encourage increased physical activity, often through structured exercise routines, goal setting, or tracking of aggregate activity such as step counts [15, 54].

A recent scoping review of older adult interaction with technologies to support their fitness [92] shows that, within this framing, interaction is typically positioned as a mechanism for increasing or sustaining activity, where movement is treated as something to be initiated, managed, or optimized. This perspective is reflected across personal informatics systems, activity trackers, gaming-based movement interfaces, and exercise-focused technologies [56, 90, 97].

While these approaches have demonstrated value in specific contexts, they typically position movement as a distinct activity—something like exercise that is undertaken in addition to daily life—rather than as a property of actions already embedded within routine practice. This raises questions about how interaction design might engage not only with increasing activity, but with how movement is understood and enacted within everyday environments, a question that becomes central in Probe 2.

### 2.2 Interactive environments and ubiquitous computing

Interactive environments and ubiquitous computing predominantly frame interaction as a **sensing–response relationship**, in which environments detect activity and act upon it [1, 98]. Ubicomp research established a vision of computation embedded invisibly within everyday settings, enabling environments to detect user activity [6], infer context [3, 98], and respond through automation or feedback [2]. Such systems have been applied across a range of domains, including energy awareness [9, 81], activity recognition [48, 80], domestic support [6] and healthcare [11, 66, 77].

In this strand, interaction is commonly framed as a relationship between sensing and response: the system observes behavior and acts upon it. While these approaches successfully embed computation within everyday environments, they tend to focus on monitoring, automation, or system-driven feedback. Less attention has been paid to how users themselves interpret their actions within these environments, or how interaction might reshape the meaning of those actions. Embedding sensing into the environment does not necessarily change how everyday activity is experienced — environments may become more responsive, but the actions performed within them remain unchanged in meaning. That is, activity is observed and acted upon, but not typically reinterpreted as interaction itself.

This vision shares ground with earlier work on calm technology [99] and ambient displays [61, 78], which proposed that well-designed systems should make information perceptible at the periphery of attention without demanding engagement. Environmental interaction draws on this sensibility — the goal is similarly to avoid intrusion and support awareness rather than compliance. However, where calm technology locates the design challenge in how a system delivers information, environmental interaction locates it in how existing actions are experienced. The system does not introduce new information into the periphery; it makes the interactional potential of actions already underway perceptible to the person performing them. More precisely: ambient and peripheral systems make information about the world continuously available at the periphery of attention. Environmental interaction makes the significance of action — what an action already is, and what it might become — continuously available to the person performing it. The persistent element is not data about behavior but the meaning of behavior in the moment of its occurrence.

### 2.3 Home-Based Health Technologies for Older Adults

Home-based health technologies for older adults predominantly frame interaction as **monitoring or guiding behavior**, positioning the individual as the subject of sensing or intervention rather than as an active interpreter of their own action.

Monitoring systems — often associated with Ambient Assisted Living (AAL) — emphasize sensing and detection, including fall detection [52, 59, 96] and activity recognition [36, 58]. These systems are predominantly passive: they operate on users rather than with them, directing outputs toward caregivers or clinicians rather than positioning the older adult as an active participant.

Intervention systems follow a different logic but share a similar framing. These typically rely on screen-based or commodity platforms — tablets [5, 39, 63], wearables [16, 30, 83, 93], or Kinect [26, 29, 68] and VR systems [8, 21, 57, 89] — to guide users through predefined exercise routines or programmes. Activity is introduced as a separate layer, requiring users to engage with dedicated interfaces, schedules, or programmes. In this sense, movement remains something to engage with through systems, rather than something already present in everyday action.

Across both trajectories, the older adult is primarily positioned as the subject of the system — monitored, detected, or guided — rather than as an agent whose existing actions carry interactive potential. This framing motivates the approach taken in this paper.

### 2.4 Embodied and tangible interaction

Embodied and tangible interaction center interaction in **intentional engagement with physical artefacts**, foregrounding the role of the body and material **artefacts and physical action** [23, 46]. These approaches emphasize the role of the body and physical artefacts in interaction, moving beyond abstract interfaces to engage directly with material and spatial experience. In tangible interaction, physical objects are designed as interfaces through which users manipulate and explore digital information [24, 38, 42, 50]. Embodied interaction shifted HCI's attention from interfaces to action situated within the world [25, 51, 55], emphasizing how meaning emerges through embodied engagement with people, artefacts [24], and environments [13, 88]. Environmental interaction builds on this foundation but draws attention to a different question: not how meaning emerges through action, but how the capability-bearing significance of actions already underway becomes recognizable and available for engagement.

More recent work has extended this focus through novel forms of physical interaction, including force-based input and interaction with deformable or movable surfaces [69–71]. Across these approaches, interaction is typically centered on intentional engagement with designed artefacts or interfaces, even when those artefacts are designed to blend closely with everyday contexts. Some work in tangible and embodied interaction has explored how interaction can be closely integrated with everyday objects and settings—for example, through shape-changing artefacts encountered in public or domestic contexts, where people make sense of a familiar object behaving differently (e.g., the coMotion shape-changing bench [35]). However, even in these cases, interaction remains centered on engagement with a **designed artefact**—the object's behavior is what invites interpretation and response. Environmental interaction differs in locus: it shifts the site of interaction from the artefact to the **action itself**, treating everyday actions (e.g., standing, reaching, lifting) as the interactional substrate, rather than as a means of engaging with a newly interactive object. That is, users act upon objects or interact with systems specifically created to support interaction.

In contrast, everyday environments are not typically designed as interaction artefacts, and the actions performed within them are not ordinarily framed *as* interaction. While some work has explored how everyday objects can subtly support interaction, these approaches still rely on designed artefacts as the primary locus of that interaction, rather than treating the action itself as the interactional site.

As a result, while embodied and tangible interaction provide important insights into the role of the body, they do not generally address how **existing, unmodified actions within everyday life might themselves become sites of interaction**.

### 2.5 Probes and exploratory design methods

Probes have been widely adopted in HCI as a means of exploring experience and opening up new design possibilities in situ. Originally introduced as cultural probes [31], the approach explicitly embraced ambiguity, incompleteness, and inspirational "fragmentary" returns as alternatives to systematic data collection or hypothesis-driven evaluation. This position was further developed by treating ambiguity as a productive design resource [32], emphasizing interpretation and personal meaning-making over clarity and optimization.

Subsequent work extended cultural probes toward more structured engagements with systems in use – the introduction of **technology probes** [43]. Technology probes reframed probes as working systems deployed in real-world contexts to simultaneously support use, inspire design, and generate empirical insight [43, 75], establishing an important bridge between exploratory design and in-situ study.

More recent work has continued to extend probe methodologies across domains. In child–computer interaction, probes have been adapted to support participatory and co-design processes that foreground children's perspectives and lived experiences [20, 44]. In HCI4D contexts, probes have been used to access situated cultural knowledge and support engagement across contexts where researchers are not embedded participants [14, 22, 104], highlighting their value in surfacing interpretation, positionality, and local meaning-making. Work with older adults has drawn on both cultural and technology probes to access how people understand and relate to their own capability, independence, and everyday environments — revealing beliefs and orientations that direct questioning or performance-based methods tend to miss [34, 37, 49, 65, 100]. Recent work on **data probes** further extends this space by using interactive artefacts to support reflection, interpretation, and dialogue around complex systems and data-driven practices [105].

Across these strands, probes provide access to situated, lived experience that is difficult to capture through controlled or purely observational methods, while foregrounding interpretation and meaning-making rather than measurement.

In this paper, we draw on probes not only as generative design tools, but as a **method for investigating an emerging interaction space**. Environmental interaction unfolds through ongoing activity, interpretation, and integration within everyday environments, and therefore does not align readily with conventional approaches that privilege discrete tasks or predefined outcomes. This is particularly true in domestic contexts with older adults, where prior probe work has shown that the most significant findings often concern how people understand the significance of their own actions — not how efficiently they perform them [34, 65]. We build most directly on the tradition of technology probes in using in-situ deployments as both functioning systems and sites of inquiry, but extend this further by focusing on how interaction is **recognized, interpreted, and enacted in context**. Deployments are treated not as evaluations of a fixed system, but as instruments for revealing how everyday actions are reinterpreted as interaction. In doing so, we shift the unit of analysis from artefacts or systems to the **environment as a site of interaction**, using probe-based deployments to map the conditions under which this form of interaction recognized, interpreted, expressed, or extended.

### 2.6 Summary: Reinterpreting everyday action as interaction

Across these four bodies of work, a consistent assumption runs beneath the surface: that interaction is something *added* to everyday life, not something already present within it. Movement is framed as an activity to be initiated or increased; environments as sites of sensing and automated response; older adults as subjects of monitoring or guidance; and interaction itself as centered on intentional engagement with designed artefacts. In each case, everyday action, sitting, standing, reaching, lifting, appears as context or raw material for interaction, rather than as interaction in its own right.

This assumption is not a flaw in any individual strand; it reflects a coherent and productive set of design priorities. But it also marks a consistent boundary. Across these traditions, relatively little attention has been paid to how actions already embedded in routine life might be reinterpreted, extended, and supported *as* sites of interaction, not prompted, tracked, or augmented from outside, but recognized as carrying interactional potential in themselves. More fundamentally, existing approaches largely assume that people already recognize the significance of their everyday actions. The possibility that such recognition may itself require design support remains comparatively underexplored.

Environmental interaction occupies this space. It does not replace existing approaches but addresses something each leaves largely unexamined: how the meaning of an action already underway can be made perceptible to the person performing it, and how that action might be optionally extended within the environment in which it occurs. This is a different design challenge from delivering prompts, embedding sensing, or designing artefacts, and it requires a different methodology.

Because environmental interaction unfolds through ongoing activity, interpretation, and integration rather than discrete tasks or predefined outcomes, conventional evaluation approaches are poorly matched to it. The probe-based tradition, particularly technology probes, offers a better fit: deployments that function simultaneously as systems in use and instruments of inquiry, oriented toward how interaction is recognized, taken up, and sustained in context. The serial probe approach adopted in this paper extends this tradition further, using a sequence of deployments to map how environmental interaction is constituted across different configurations of system, environment, and participant circumstance.

## 3 APPROACH AND DESIGN RATIONALE: FROM MAINTAINING INDEPENDENCE TO ENVIRONMENTAL INTERACTION

The work presented in this paper builds on several years of exploratory co-design and prototyping with older adults around health, movement, and independence. These engagements were semi-structured and generative, focusing on participants' everyday experiences rather than formal evaluation; the empirical findings reported here are drawn from the probe studies that follow.

Across these engagements, a consistent concern emerged. Participants were not primarily motivated by a desire to "exercise more," but by maintaining independence in everyday life: the ability to stand up safely, reach for objects, move confidently, and avoid reliance on others for personal care. These concerns were grounded in lived experience and anticipated loss, with physical capability closely tied to autonomy, dignity, and quality of life. While the importance of independence in later life is well established [67, 94], less attention has been paid to how this translates into the design of everyday interaction within domestic environments.

### 3.1 Revealing the link between strength and everyday action

Within this framing of independence, a key observation emerged. Many of the actions participants cared about—standing, lifting, stabilizing—are underpinned by strength, balance, and mobility, yet were not typically recognized as forms of strength practice. When introduced to the idea that everyday actions could already constitute strength work, participants often reacted with surprise:

*"Oh—you're kidding… I never thought about it that way… I can do that!"*

Simple reframings—such as repeating a sit-to-stand movement ("do it twice")—were immediately intelligible and often experienced as achievable and meaningful. Rather than introducing a new activity, this reframing enabled participants to recognize what they were already doing and consider how it might be extended.

This also resonated with existing strategies. For example, participants described adapting activities (such as using an exercise bike when outdoor walking was not possible) to maintain capability, suggesting that movement could already be translated into different forms when tied to meaningful goals. However, the interpretation of everyday action as structured strength practice is not typically made explicit in either daily life or system design.

### 3.2 Not new behavior, but recognizing and extending common action

Building on this reframing, participants did not identify the challenge as adopting new behaviors, but as:

- recognizing what they were already doing
- remembering to extend those actions
- understanding how to do so safely and effectively

Participants consistently identified the need for:

- **Reminders** — to prompt action in context, given acknowledged patterns of forgetting
- **Guides** — to indicate how actions should be performed or extended

Importantly, these were not requests for structured exercise programmes or new routines, but for support **embedded within ongoing activity**. Participants did not frame this in terms of modifying behavior, but in **recognizing and extending actions already embedded in daily life**. This framing establishes a design direction: not increasing activity per se, but making movement legible, actionable, and extendable within existing routines.

### 3.3 Enlivening the environment as a site of interaction

These observations point to a shift in how interaction is framed. Rather than introducing new activities or interfaces, interaction can be understood as emerging through the reinterpretation and extension of actions already embedded within everyday environments.

From this perspective, the role of the environment is not simply to host activity, but to support interaction by: *revealing* the movement properties of everyday actions, *supporting* their extension through reminders and guidance, *maintaining* the functional capabilities those actions enable. For example, pressing up from a chair using one's arms can act as a proxy for movements required to exit a bath. In this way, everyday environments provide opportunities to sustain capability in situ, without requiring separate settings or activities.

These observations established a design direction. The challenge was not to introduce new behaviors but to support recognition and extension of actions already present. Our initial instrumented prototypes confirmed that the signals needed to detect such actions — sit-stand transitions, grip, balance shifts — could be reliably captured. This preliminary work established the technical foundations for the refined system which was deployed in our probes, described in Section 4, below, to study how such interaction is understood and taken up in practice.

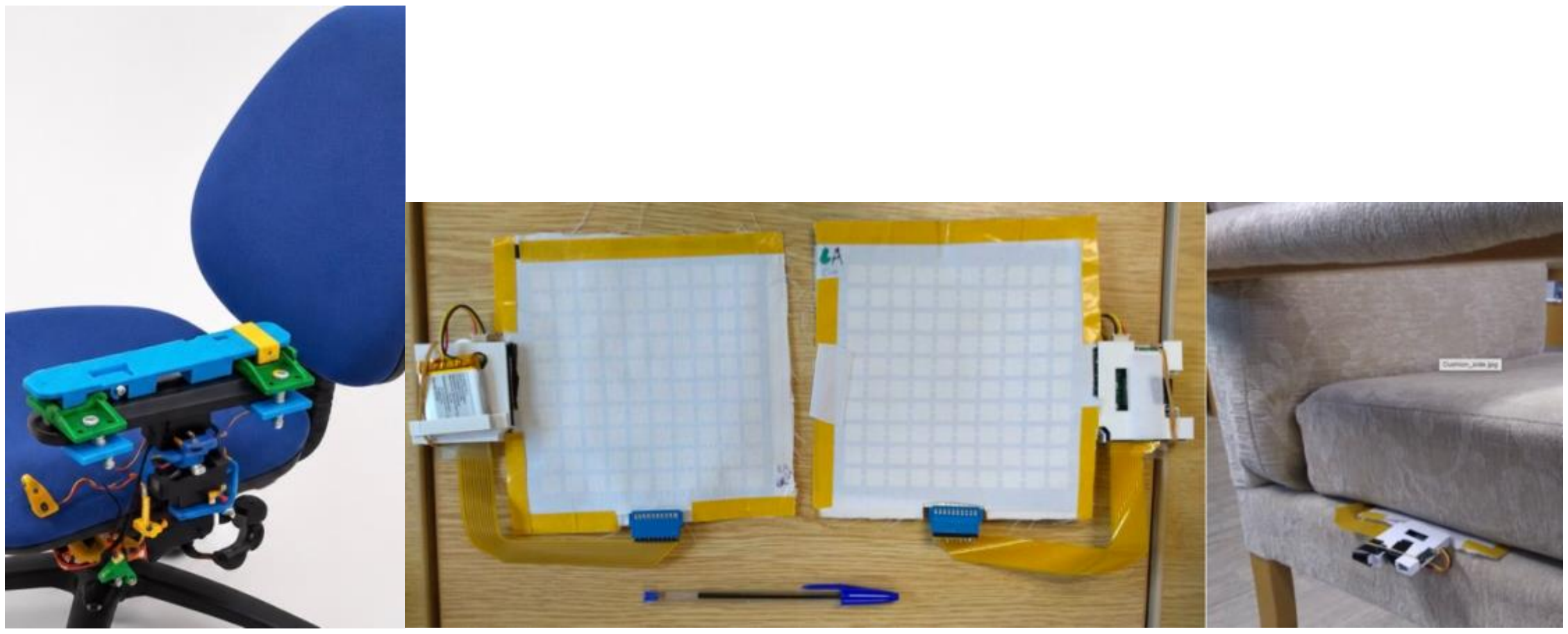

Figure 2**:** Example of early hard-printed prototype parts (our FrankenChair arm press up detector), left, to softer, flexible velostat sensors and custom micro controllers to detect sit/stands, towards final evolution as per Figure 3.

## 4 METHODS

We investigate environmental interaction through a series of four in-situ probe deployments, each treating the deployment itself as an instrument of inquiry rather than a system under evaluation. The probes move from establishing technical viability in a lab-proximate setting, through examining how guidance is interpreted and organized in the home, to exploring how alternative representations of guidance might better support recognition, invitation, and presence in everyday environments. The system deployed across these probes is described below; the interaction logic specific to each probe is introduced as each probe is presented.

### 4.1 Apparatus

Building on the preliminary prototyping work described in Section 3, the deployed system consisted of sensing-enabled elements — instrumented seating, floor surfaces, and hand-held objects — coordinated through a central unit to detect everyday movement and offer optional, context-sensitive guidance. The deployed system consisted of sensing-enabled elements — instrumented seating, floor surfaces, and hand-held objects — coordinated through a central unit to detect everyday movement and offer optional, context-sensitive guidance. Rather than introducing new tasks or devices, the system detects and characterizes movement already occurring in everyday activity, capturing signals such as pressure distribution, sit-stand transitions, balance shifts, and object manipulation. These signals are processed to identify everyday actions as movement events, which can then be reflected back to participants through minimal, context-sensitive feedback supporting recognition and optional extension of those actions. The system was designed to operate without internet connectivity, reflecting participant privacy concerns, and ran continuously without requiring explicit activation (Figure 3). One element, the CanBand (Figure 4) — an augmented hand-held object with embedded LEDs — additionally supported detection of lifting and reaching actions, with simple feedback indicating movement direction or repetition (Figure 5). Full technical details are in [28]; the interaction logic is described as each probe is introduced

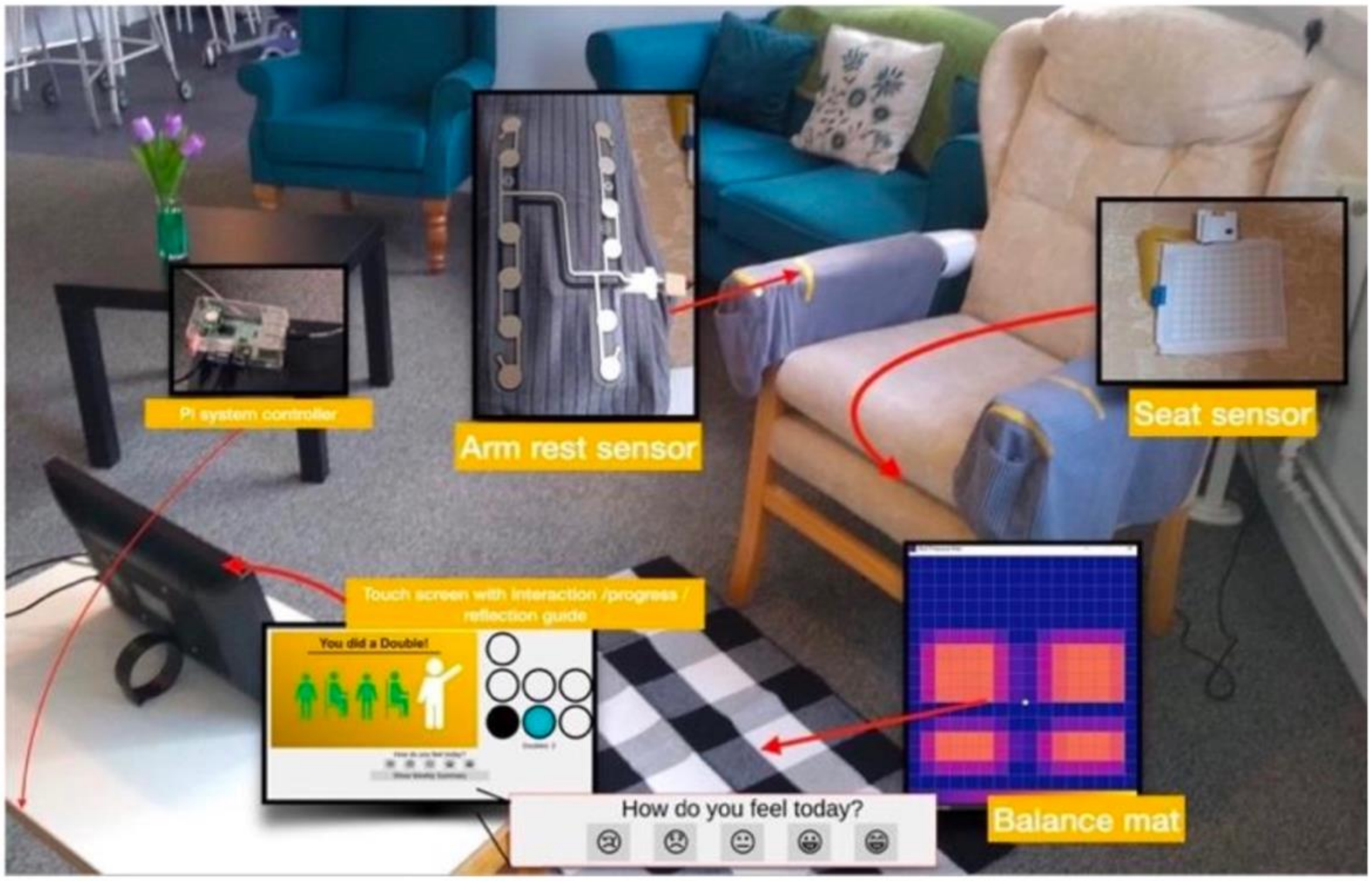

Figure **3**: In-lab setup for Probe 1, approximating a domestic environment.

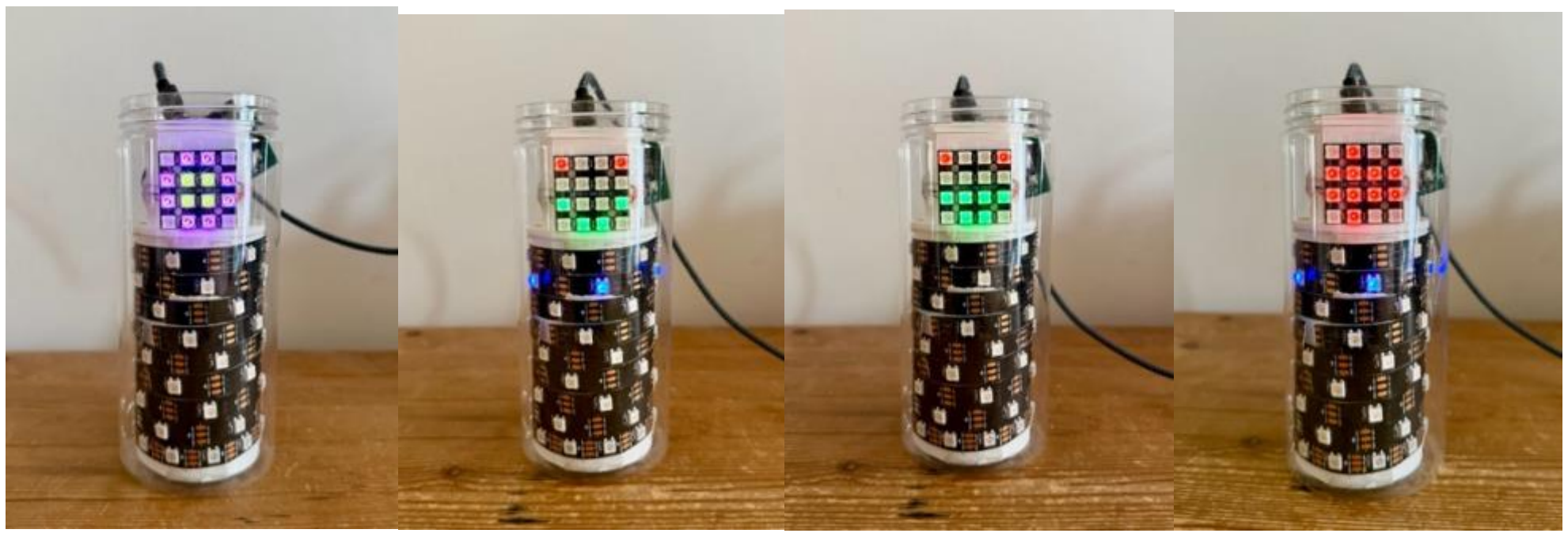

(1) (2) (3) (4)

Figure **4**: The CanBand - (1) the LED matrix flashes as an invitation to reach and grab the device, (2) a smiley face representation to inform a single reach-lower-putback movement was completed, (3) a different smiley face if a “double” was complete and (4) arrow display to show the direction to move the CanBand.

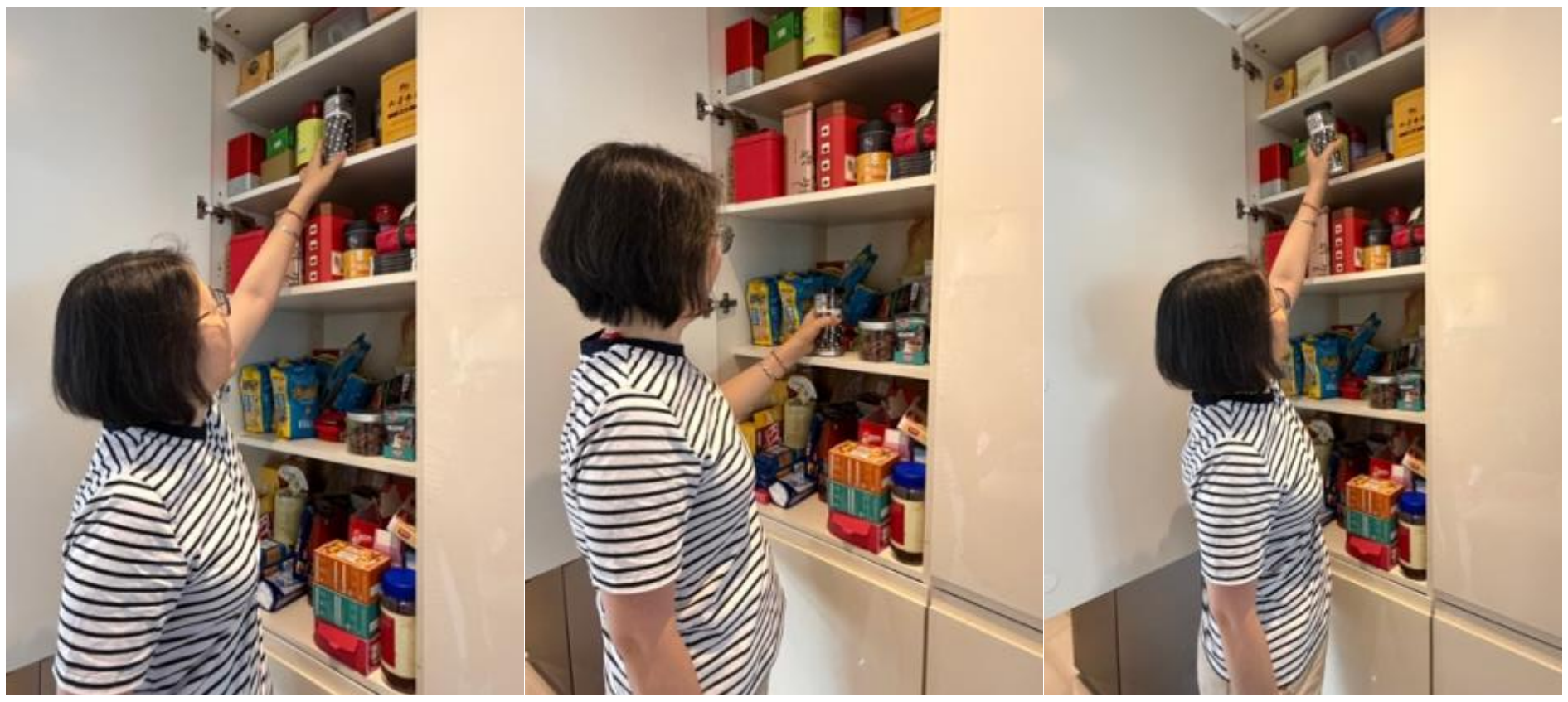

(1) (2) (3)

Figure **5**: An example of how the CanBand fits in a home environment: when the user opens the cupboard, the CanBand will flash to invite the user to (1) reach and grab it, (2) lower it and (3) place it back in the original position to complete one rep.

### 4.2 Methodology: Serial Probes

To investigate this design space, we adopt a probe-based approach grounded in the tradition of cultural [31] and technology probes [43]. This establishes a design direction; the probes are deployed in series to examine how users make sense of and engage with systems designed to cue environmental interaction.

We frame this approach as ***serial probes***: a sequence of studies in which each probe is defined not primarily by a fixed deployment format, but by its epistemic role in exploring an emerging interaction space. Rather than refining toward a single solution, each probe functions as an instrument of inquiry, reconfiguring the questions raised by the previous one.

Across the four probes presented below, this progression moves from establishing technical instantiation (Probe 1), to examining how guidance is interpreted and taken up in everyday life (Probes 2 and 3), and finally to exploring how alternative representations reshape interaction's perceived fit, presence, and meaning within the home (Probe 4). Observations—including breakdowns, reinterpretations, and adaptations—directly inform both subsequent system configurations and the questions pursued, allowing the inquiry itself to evolve.

This approach is particularly suited to environmental interaction, where interaction unfolds through ongoing activity and interpretation rather than discrete tasks, and where outcomes are not fully specified in advance. Here, probes are used not to evaluate a fixed interaction technique, but to explore how everyday actions are recognized, reinterpreted, and taken up as interaction within the routines and contexts in which they occur.

In this way, the probes operate serially to trace how environmental interaction is recognized, shaped, and integrated over time and across configurations of environment, system behavior, and participant context. Taken together, they map the conditions under which such interaction becomes meaningful, actionable, or is not taken up.

In this sense, our approach extends the role of technology probes from exploring systems in use to exploring how interaction itself is constituted within everyday environments.

Across probes, inclusion and exclusion criteria were designed to ensure participants could safely engage with repeated sit–stand and lifting activities, while reflecting the everyday contexts in which such interaction would occur.

### 4.3 Participants and Recruitment (Probes 1–3)

Participants for Probes 1–3 were recruited through outreach to local community centers, care homes, university mailing lists, and poster advertisements. This approach aimed to engage individuals living independently in the community who were interested in maintaining or improving everyday movement and strength.

#### *4.3.1 Rationale and Evolution of Criteria*

Participant selection evolved across the probe sequence as part of the inquiry itself.

Probe 1 prioritized broad accessibility, applying flexible inclusion criteria (age 65+) to support exploratory testing in a controlled environment and assess technical feasibility across a range of individuals.

Probes 2 and 3 applied more consistent inclusion criteria — adults aged 50 or over, English-speaking, with sufficient physical capacity for repeated sit-stand movements — to support safe and sustained in-home deployment. Observations from Probe 2 indicated that participants who were already active often interpreted the interaction as a task to complete, rather than an opportunity embedded in ongoing activity.

In response, Probe 3 focused on participants who identified as less active or who expressed a desire to increase movement. This shift enabled examination of how environmental interaction is taken up when there is greater alignment between system intent and participant motivation.

As such, participant selection across Probes 1–3 is not treated as a fixed population, but as part of the probe-based inquiry, allowing variation in uptake to be explored across different contexts of activity, readiness, and everyday practice. All participants provided informed consent and were compensated in accordance with institutional guidelines..

#### *4.3.2 Participant Overview*

Across Probes 1–3, a total of 13 participants took part in the study programmed, as per Table 1 and **Error! Reference source not found.**.

Table 1: The number of participants in Probe 1-3 and the characteristics of participants for each probe study.

| Probe | N | Context | Participant characteristics (summary) |
|---|---|---|---|
| Probe 1 | 7 | In-lab | Broad exploratory sample; mixed activity levels |
| Probe 2 | 3 | In-home (2 weeks) | Participants already engaged in regular daily activity |
| Probe 3 | 3 | In-home (4 weeks) | Participants identifying as less active or seeking to increase movement |

Table **2**: Specific demographics of participants in Probe 2 and 3. In Probe 3, P2 and P3 were co-habitants in the same home.

| Probe | Participant | Gender | Age |
|---|---|---|---|
| Probe 2 | P1 | Male | 73 |
| Probe 2 | P2 | Female | 90 |
| Probe 2 | P3 | Female | 71 |
| Probe 3 | P1 | Male | 56 |
| Probe 3 | P2 | Female | 76 |
| Probe 3 | P3 | Male | 79 |

Technology probe studies routinely work with smaller groups not least because the goal is depth of situated insight rather than statistical generalizability. We take our inspiration from Hutchinson et al. [43]'s original technology probes work who tested probes with four families: eight members (US), four members (Sweden) and two separate families of siblings (France). Subsequent deployments have used similarly small numbers: Cyrato et al. [19]'s Kinect study with four participants and Stawarz et al. [90]'s in-home study with five participants.

#### *4.3.3 Data Capture and Ethics*

Across Probes 1–3, data were collected through a combination of:

- For all three probes, observational notes during deployment and interaction
- For all three probes, post-session or post-deployment user surveys
- For probe 1, in session semi-structured interviews
- For probes 2 and 3, in situ system logging including the number, time of day, and speed of single and double sit-stand movements per day and the number of days daily goals were met
- For probes 2 and 3, pre- and post- health state and strength measurements

These methods were used not to evaluate predefined outcomes, but to support interpretation of how participants understood, engaged with, and incorporated interaction into everyday activity. Analysis of data from Probe 2 and Probe 3 was structured in relation to the design considerations developed in [91], which synthesizes older-adult-specific technology acceptance factors from the literature. This provided an explicit analytic framework for examining how participants' engagement aligned with or diverged from expressed needs and preferences across dimensions including compatibility with lifestyle, dignity and independence, and emotional engagement. Qualitative data from observations and surveys were interpreted thematically within this framework, with system logs providing behavioral corroboration of reported experience. Probe 4 sessions were transcribed and analyzed thematically, focusing on how participants interpreted, evaluated, and situated alternative representations within the context of their own homes and identities.
All procedures were conducted with institutional ethical approval [ethics numbers/institutions blinded for review].

#### *4.3.4 Probe 4: Participant Approach*

Probe 4 adopts a distinct participant and methodological approach, reflecting its role as a comparative and reflective exploration of alternative representations. Participants were recruited through community centers, care homes, NHS partners, and local outreach channels. Eligibility criteria were broadened (age 50+) to include individuals at earlier stages of considering independence and everyday capability, supporting reflection across a wider trajectory of ageing (Table 3).

Probe 4 combined one-to-one interviews and group sessions of varying sizes. This flexible format was used to create multiple opportunities for participation across different contexts and availability, enabling engagement with a broader range of participants. While session formats varied, consistent patterns emerged across groups, supporting comparative interpretation of responses (Figure **6**).

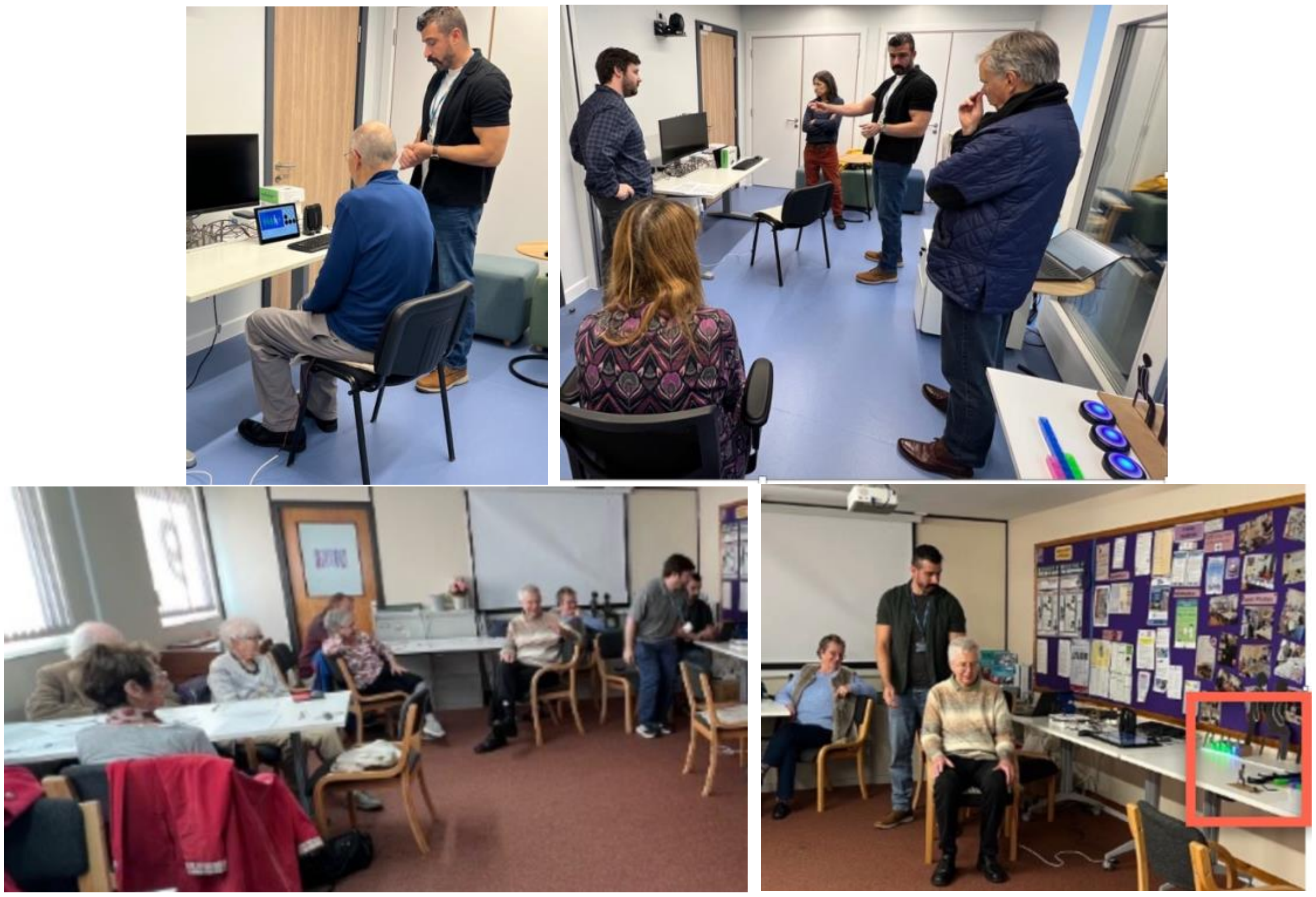

Figure **6**: Probe 4 focus groups: Top row shows in-lab one-on-one demonstration and discussions. Bottom row shows community group demonstrations with volunteer community member.

A total of 25 adults participated over six sessions: three one-on-one interviews, one three-person small group session one five-person session, each held at our university, and one 14-person session held at a local community center. Participants were given a gift card voucher for their participation. Table 3 shows the age groups of the participants who took part in the Probe 4 study.

Table 3: The age range of the participants who took part in Probe 4.

| **Age Range** | **No. of Participant** |
|---|---|
| 50 – 59 | 2 |
| 60 - 69 | 8 |
| 70 - 79 | 11 |
| 80 - 89 | 3 |
| 90+ | 1 |

### 4.4 Probe 1: Establishing technical viability of in situ movement guidance

**Purpose** As the first probe in this serial sequence, Probe 1 addresses a foundational question: can environmental interaction be materially instantiated in practice? Specifically, can everyday actions be reliably detected and coupled with intelligible, in situ guidance aligned to those actions?

This probe focuses on two forms of everyday activity:

- **sit–stand movement**, representing a frequent, whole-body transition embedded in daily routines (Figure 3).
- **object-based movement**, explored through the CanBand as a proxy for lifting and reaching actions (Figure 5).

The aim is to establish whether such actions can be recognized and extended through simple prompts (e.g. "do it twice"), and whether this coupling of action and guidance is intelligible to participants in real time. While this overlaps with a conventional HCI concern with system functionality, its role here is broader: to establish whether environmental interaction can be instantiated sufficiently to become available for further inquiry.

**Deployment**

The system (Section 4.1) was deployed within an in-lab environment designed to approximate everyday domestic conditions (Figure **3**). Participants interacted with instrumented seating (seat cushion and armrests), floor-based sensing, and a central guide display.

The primary interaction explored was **sit–stand movement**. When a transition was detected, the system provided brief, optional guidance sequences—for example, inviting repetition ("do it twice"). Variants included arm-assisted and hands-free standing, as well as simple weight-shifting behaviors, allowing exploration of how familiar movements might support strength, balance, and coordination.

In parallel, the system included an exploratory **object-based interaction** using the CanBand, enabling simple up–down lifting and placement sequences. This provided a contrasting mode of interaction through tangible manipulation.

**What happened**

Across participants and sessions, the system was able to detect sit–stand transitions and respond with guidance aligned to participants' actions. Participants were able to follow prompts presented on the display, including repetition cues such as "do it twice," and demonstrated understanding of the relationship between their movement and system feedback during in-session observation.

Direct observation indicated that participants were comfortable performing the sit–stand interaction, with 6 of 7 participants freely engaging in repeated sit–stand movements over the session, including both single and repeated ("double") actions.

A practical constraint emerged relating to sensor placement and participant variability. For one participant, sit–stand transitions were not reliably detected due to differences in seating position, which affected pressure distribution on the instrumented cushion. Adjustments to sensor placement resolved this issue and informed refinements for subsequent probes.

Participants were also able to interpret variations in guidance, including verbal instructions to use arm support or attempt unsupported standing. These interactions aligned closely with movements already present in everyday activity.

In contrast to the largely structured sit–stand interaction, the CanBand elicited more exploratory engagement. Participants frequently repeated and varied lifting and placement actions to observe how the system responded—for example, changes in feedback patterns or rewards such as the "happy face" indicator. Several participants remarked that this interaction was enjoyable and continued beyond the initially prompted sequence.

This suggests that, even at this early stage, interaction was not limited to task completion but could also involve curiosity-driven exploration, depending on how action and feedback were coupled. At the same time, observations indicated that response timing in the CanBand interaction was not always sufficiently fast to clearly support interpretation, highlighting limitations in the current implementation.

**What this Suggests**

From a conventional HCI perspective, this probe establishes technical viability: the system can detect everyday movement, provide intelligible guidance, and support interaction across both whole-body and object-based actions. More significantly for the inquiry, it establishes that environmental interaction can be materially realized and encountered in practice — that everyday actions, without modification, can become sites of recognizable interaction. Participants were able to engage with guidance as an invitation within ongoing activity, and early observations suggested that this engagement varied depending on how action and feedback were coupled: curiosity-driven exploration emerged alongside task-oriented responses, indicating that the interactional quality of the system — not just its functionality — was already shaping uptake. At the same time, constraints emerged: response timing in the CanBand interaction was not always sufficiently fast to support clear interpretation, and the lab setting meant that questions of how such guidance would be taken up, organized, or sustained in everyday domestic life remained open. These constraints and questions shape the subsequent probes.

As the first probe in this sequence, its contribution is therefore not only to show that the system can function, but to establish that environmental interaction can be materially realized and encountered in practice, while surfacing early constraints and differences in engagement that shape the next probes.

How environmental interaction is taken up, organized, and sustained beyond the lab — and how guidance is understood within the rhythms of everyday domestic life — motivates the subsequent probes..

### 4.5 Probe 2: Everyday deployment – interpreting guidance in practice

**Purpose**

As the second probe in this serial sequence, Probe 2 examines how environmental interaction is interpreted when moved from the controlled setting of the lab into participants' everyday domestic environments over a two-week period. Building on Probe 1, which established that environmental interaction could be instantiated in practice, this probe asks how that guidance is understood and organized in daily life.

The central question here is not simply whether the system continues to function reliably in the home, but whether the "do it twice" interaction is taken up as an embedded extension of ongoing activity or instead interpreted and enacted in some other way.

**Deployment**

The system was deployed in participants' homes for two weeks that included the sit-stand detection, the "do it twice" reminder and guide screen, and the CanBand. Participants were introduced to the system during in-person setup, where the intended interaction was explained and demonstrated. Following Probe 1, the system was also refined to improve robustness in home use, including updates to sit–stand detection through changes to pressure sensing, sensor positioning in the chair, and the detection algorithm.

**What happened**

In practice, the system functioned well for sustained in-home use, and participants were able to perform the movements required by the interaction. Use of the can-band component however, highlighted some practical constraints. System logs showed that participants interacted with the device daily, but battery life was significantly shorter than in prior testing. Replacing the batteries with a larger pack addressed this issue, but introduced additional weight, making the device less comfortable for repeated use. While this did not prevent engagement, it influenced how the component was incorporated into everyday activity. With the chair deployment, system logs showed that participants met, or came close to meeting, their daily sit–stand goals across the deployment. However, the guidance was not consistently taken up in the intended way. Participants could understand the prompt and complete the required repetitions, but interaction was interpreted differently across participants.

For two participants (acquaintances), the interaction was largely treated as something to get done, rather than something to weave into everyday movement. System logs showed that their "double" sit–stand repetitions were frequently bunched together in the first few hours of the morning, suggesting that the interaction had been reorganized into a bounded task rather than distributed across ongoing activity. By contrast, the third participant described the system more positively, framing it as a useful opportunity within everyday life: "*It was an extra activity, so always helpful…it reminded me to stand and move*." In this case, the interaction was not described as a chore-like task to complete, but as creating positive opportunities for movement within everyday activity. Probe 2 therefore establishes an important distinction: the system could be used successfully, but successful use did not necessarily correspond to the intended model of environmental interaction.

At this stage, the probe does not resolve why this occurred. Rather, it shows that environmental interaction cannot be assessed solely in terms of prompt delivery, compliance, or target achievement. It must also be understood in terms of how guidance is interpreted and situated within everyday life. As the second probe in this sequence, Probe 2 therefore surfaces a central question for the remainder of the inquiry: how can reminders and guides be represented, introduced, and positioned so that they are encountered as part of ongoing action rather than as another task to complete? This question shapes the subsequent probes, which explore variation in uptake across participants and contexts (Probe 3) and the representational forms through which such guidance may better fit, engage, and persist within the home (Probe 4).

**Transition: From Probe 2 to subsequent probes**

Probe 2 did not resolve into a single design direction. Instead it revealed that the inquiry had opened in two directions at once.

The first concerned the system itself: the interaction was usable, but participants had reorganized it as a bounded task rather than something encountered within ongoing activity. This raised a practical question about whether changes to onboarding, interface design, and system configuration could better support extension within the moment of action — a question explored through Probe 3.

The second direction concerned representation and fit. Probe 2 brought the inquiry back to something participants had raised in the original co-design work: the need for reminders and guides that fit within everyday life. Where that need had initially seemed like a design specification, Probe 2 revealed it as a deeper question about how a reminder looks, where it sits, and what it signals about the person who lives with it. This thread was carried into Probe 4, which returned to reminders and guides not as a step backward but as a spiral: the same question, now informed by what the preceding probes had revealed about uptake, interpretation, and fit.

Although these trajectories were developed in parallel — design work for Probe 4 proceeding while Probe 3 was running in homes — they were not independent. Probe 4 sessions waited for Probe 3 responses, so that what sustained in-home deployment revealed could inform how alternative representations were explored.

### 4.6 Probe 3: Sustained engagement – variation across people and contexts

**Purpose**

Probe 3 follows the first trajectory identified after Probe 2: examining how environmental interaction is taken up across people and contexts over time. Building on Probe 2—which showed that guidance could be used but was often reorganized as a bounded task—this probe asks how engagement with the refined system varies in sustained everyday use. The CanBand component used in Probe 2 was not included in Probe 3. Addressing the practical constraints observed in its use (battery life and weight) required additional development time. Given the parallel progression of subsequent probes, we chose to focus Probe 3 on variation in uptake of the sit–stand interaction, with Probe 4 returning to questions of system form and integration. The aim for Probe 3, therefore, focused on whether refinements to onboarding and system configuration support interaction as an extension of ongoing activity—and how such interaction is differentially taken up across participants and domestic contexts.

**Deployment**

The system was deployed in participants' homes over an extended four-week period. Following Probe 2, the system was refined through design workshops, including removal of the floor mat, addition of a sleep mode, configuration controls for visibility of daily goals, and improved onboarding materials. These changes were intended to make the system easier to live with and to support distributed use throughout the day rather than session-based interaction.

The core interaction remained optional sit–stand guidance embedded in everyday activity, supporting extension of actions already in progress. Participant selection also shifted relative to Probe 2: participants were recruited who identified as less active than they would like to be and who expressed interest in moving more.

**What happened**

Engagement with the system varied markedly across participants, revealing differences not only in degree of use but in how the interaction was taken up in practice.

One participant demonstrated strong and sustained engagement across the deployment and described the intervention as "*overall … a very motivating and enjoyable experience*." In contrast, a cohabiting couple showed minimal sustained engagement over the same period, reporting that they had "*continued to live normally, standing or sitting when [they] wanted to*." At the same time, these participants' retrospective accounts do not fully align with observed behavior: system logs indicated that use had largely ceased after the first week, suggesting a divergence between reported reasoning and actual uptake.

Taken together, these cases indicate that even when the system is usable and refined, uptake is not uniform but varies across participants and domestic situations.

**What this Suggests**

Probe 3 shows that improving onboarding and simplifying the system are not sufficient to ensure consistent uptake of interaction within ongoing activity.

Even under similar system conditions, participants differed not only in how much they engaged, but in whether the interaction was taken up at all. This probe does not resolve the causes of this variation, but shows that it cannot be explained solely at the level of system configuration or instruction.

Instead, the findings suggest that uptake depends on factors beyond the immediate interaction design, including how participants recognize, value, and incorporate such interaction within their existing routines and circumstances.

These issues are developed further in subsequent probes and synthesized in Section 6, particularly in relation to recognition before extension and the role of context in shaping uptake.

### 4.7 Probe 4: Reflective exploration – representation, meaning, and fit in the environment

**Purpose**

Probe 4 follows the second trajectory identified after Probe 2, returning to the question of reminders and guides to examine how they are represented within the home.

While earlier probes showed that guidance could be delivered and acted on, they also revealed that interaction was frequently reorganized as a task, and that uptake varied across participants and contexts. Probe 4 therefore addresses a distinct question: how do different representations of guidance shape whether environmental interaction is recognized, encountered, and engaged with as part of everyday life?

Rather than refining a single system, this probe adopts a comparative approach, examining how alternative forms of representation influence interpretation, fit, and engagement within domestic environments.

**Deployment**

Instead of deploying a single system in situ, this probe used focus groups and interviews to explore contrasting representations of reminders and guides.

Participants were introduced to three forms of guidance derived from earlier probes and subsequent iterations:

1. **Screen-based guidance**, presenting sit–stand prompts as visual instructions (Figure **7**).

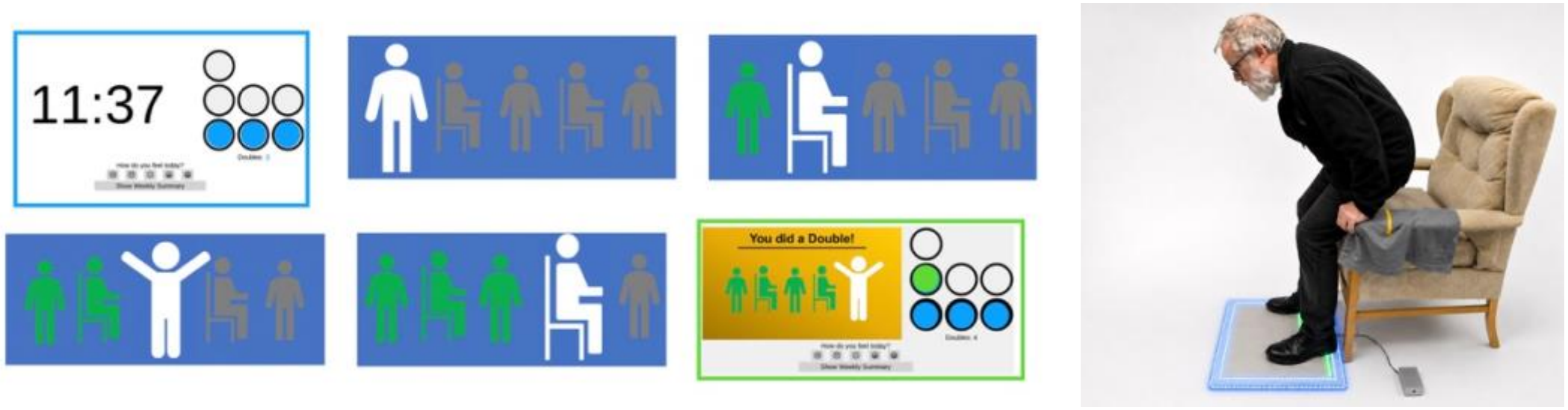

Figure **7**: the sit-stand sequence as reminder and guide in our screen display. The home screen displays the current time and the number of "do it twice" sit-stand+sit-stand's completed so far. The system is activated when it detects the user is standing (first white figure).

The system then guides the user by informing the current state (figure in white) and invites the user to do the next action in the sequence. Completed actions in the sequence are colored in green.

2. **A motorized sit/stand figure ("Mr. Z")**, modeling movement through animated posture (Figure **8**)

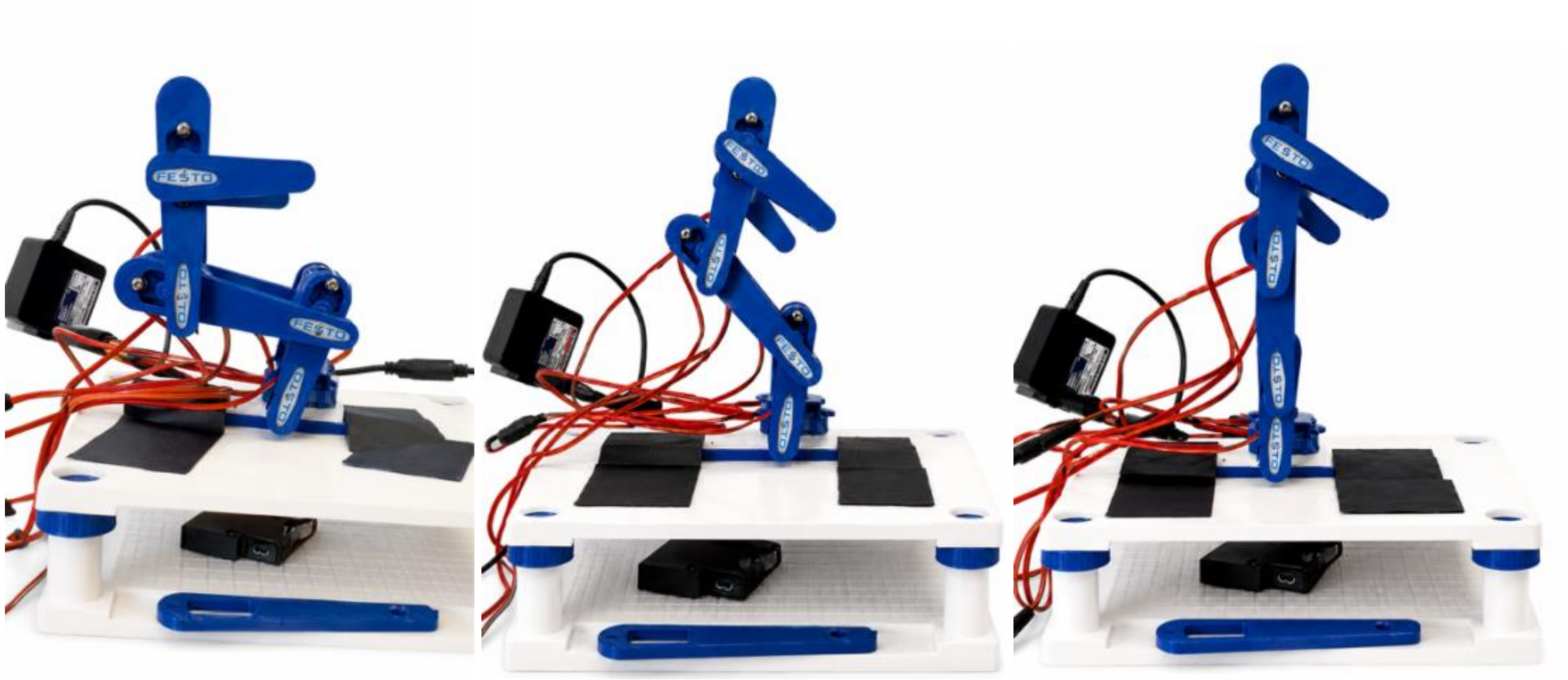

Figure **8**: Mr. Z prototype showing sit to stand sequence (from left to right).

3. **Physicalised and ambient representations**, including static figures used to explore size (Figure **9**), and similar figures where light-augmented sequences (Figure **10**), and sound-supported the sit/stand cues *[uploaded video for review]*.

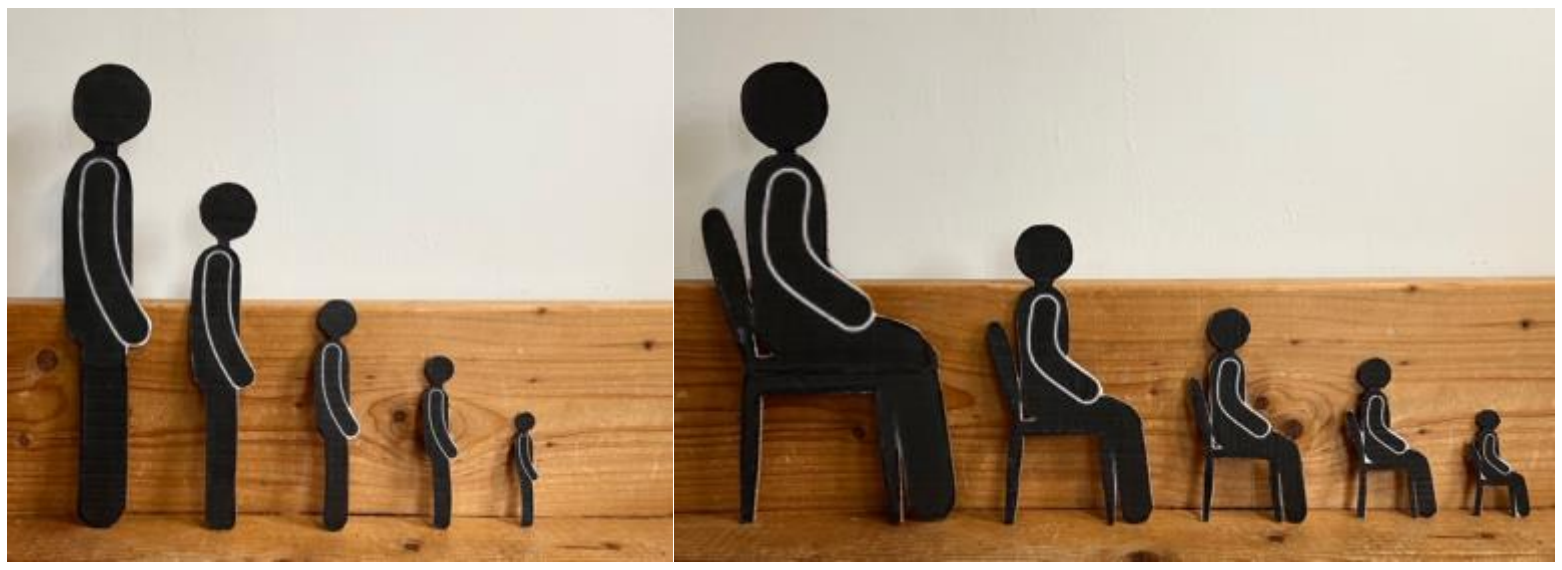

Figure **9**: The height of the standing (44, 33, 22, 16.5 11 cm) and sitting (40, 30, 20, 15 10 cm) figures.

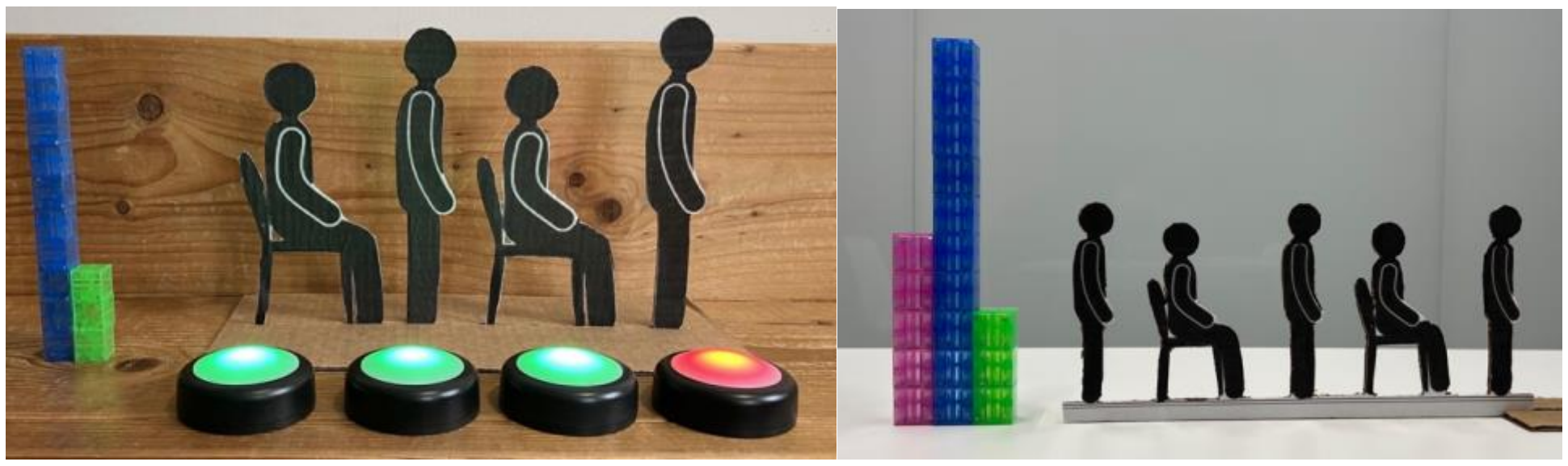

Figure **10**: Reminder and guide prototype with large light display as well as physical data visualization (left). Smaller figures with physical doubles goals, achieved singles, achieved doubles represented (right).

These figures were presented in multiple configurations and scales, allowing participants to consider how each might be placed and experienced within their own homes.
Discussions focused on how different representations influenced:

- interpretation of the interaction
- fit within the home
- and how guidance might support or disrupt everyday activity

**What happened**

Participant responses were shaped strongly by representational form — not just in terms of clarity or usability, but in relation to how each form would exist within everyday life.

Screen-based guidance was immediately legible but consistently described as poorly integrated: participants already had screens, did not want another, and experienced the prompt as something that appeared, demanded attention, and disappeared. This episodic quality — interaction as an event — ran directly against the goal of fitting within ongoing activity. The motorized figure, "Mr. Z", (Figure **8**) raised a different problem: rather than leading action, it was frequently perceived as reacting to it, and its timing, scale, and motor noise affected whether it read as guide or gadget. Together, these responses suggest that effective guidance requires not only representation of action but a clear positional role, in this case, something that invites rather than follows.

Physicalized and ambient representations shifted this dynamic. Participants responded most positively to forms augmented with light and sound, describing them as less demanding of attention, more flexible in use, and able to exist as a persistent presence rather than an episodic prompt. Participants consistently preferred guidance that functioned as invitation — "it's just there, you can do it if you want" — over anything directive or obligatory. Light- and sound-augmented forms were often described as genuinely engaging, with participants suggesting adaptations like muting sound, indicating an expectation of flexibility and personal control. This distinction — presence rather than event, invitation rather than instruction — emerged as the central organizing preference across sessions.

Participant responses to representational form extended beyond usability considerations in ways that existing frameworks for domestic technology fit do not readily account for. Artefacts that seemed functionally appropriate were rejected on grounds unrelated to function: 'that makes me feel old,' 'things that don't look like a hospital are better.' Participants were reading these objects not as tools but as statements about identity and bodily self-perception — a dimension of fit that sits outside conventional usability or compatibility frameworks. This suggests that designing for environmental fit requires attention not only to physical unobtrusiveness but to how objects signify within people's understanding of themselves and their capabilities

Attempts to physicalize activity data — representing completed repetitions as tangible objects — generated limited engagement as part of the in-situ guidance experience. However, they did prompt participants to articulate what reflective information they would want, and where: consistently, in a separate tool such as an app, accessed deliberately rather than encountered passively. This reinforces the doing/reflecting distinction while also suggesting that the appetite for reflection exists — it simply belongs in a different register.

One finding cut across all representational forms and all session formats. When participants were told that actions they already perform daily constitute meaningful strength work — that a chair sit-stand is a box squat, that pushing up from a bath is a dip, that pulling open a door is a row — the response was consistent surprise: "oh that's really neat," "I hadn't realized that," "I hadn't thought of it that way." And not least "I can do that!" This reaction appeared in one-to-one interviews, small groups, and the larger community session alike, and was not prompted — it arose each time the reframing was introduced. That the same moment of recognition continued to land as a genuine revelation across Probe 4's broader and more varied participant group suggests something important: recognizing everyday action as meaningful capability cannot be assumed, and the system as deployed did not support it. The "do it twice" invitation only makes sense once that foundation is in place. This question — how recognition itself might be actively supported — carries directly into the design considerations that follow.

**What this suggests**

Probe 4 shows that representation is not a surface question. How guidance looks, what it signals, where it sits, and how it persists within a home shapes whether environmental interaction is experienced as invitation or instruction, presence or event, opportunity or obligation. These are not usability distinctions — they are questions about how interaction is constituted within the social and material texture of everyday life. Designing reminders and guides for this space is therefore not a matter of delivering prompts clearly, but of shaping how action is made visible, meaningful, and available — in forms that fit not only the environment but the person who inhabits it.

## 5 DISCUSSION: TOWARD ENVIRONMENTAL INTERACTION IN EVERYDAY LIFE

Across these probes, we examined whether everyday action could function as a site of interaction to support movement within everyday environments are interpreted, taken up, and sustained in practice. While the system was technically robust and capable of detecting and guiding action (Probe 1), subsequent probes revealed a set of interaction challenges that extend beyond functionality, pointing toward a different framing of interaction in this domain. The purpose of these probes was not to validate a completed interaction paradigm but to investigate whether everyday action could function as a site of interaction. In doing so, the probes repeatedly surfaced a prior and unanticipated challenge: recognition. The contribution of this work is therefore not a solution to recognition, but the identification of recognition as a prerequisite for environmental interaction and a design problem that existing movement-support paradigms largely leave unaddressed.

Across the probe sequence, the inquiry returns to the participant-derived need for reminders and guides first surfaced in the formative work, but now with clearer evidence that these are not simple prompt-delivery problems: they are questions about how action is recognized, interpreted, and situated within everyday life.

Taken together, the probes suggest that **recognition of the capability-bearing significance of everyday action cannot be assumed**. Participants could successfully perform the guided actions while simultaneously interpreting them through existing models of tasks, exercise, and technology use. This finding shifts attention from the delivery of guidance to the prior challenge of recognition itself. The recognition challenge was not anticipated at the outset of the inquiry; it emerged when Probe 2 revealed that participants could successfully perform the guided actions while simultaneously assimilating them into existing task-completion models. Supporting recognition as a process, as sustainable, therefore emerges as a distinct design challenge. In this sense environmental interaction with its focus on *recognition / extension* at the site of the interaction (such as about to sit, about to answer the door, about to make tea) is not only a matter of sensing and feedback, but of **how actions are represented, interpreted, and integrated within everyday life over time (**Figure 11). In this section, we synthesize these observations into a set of considerations for designing interaction in this space.

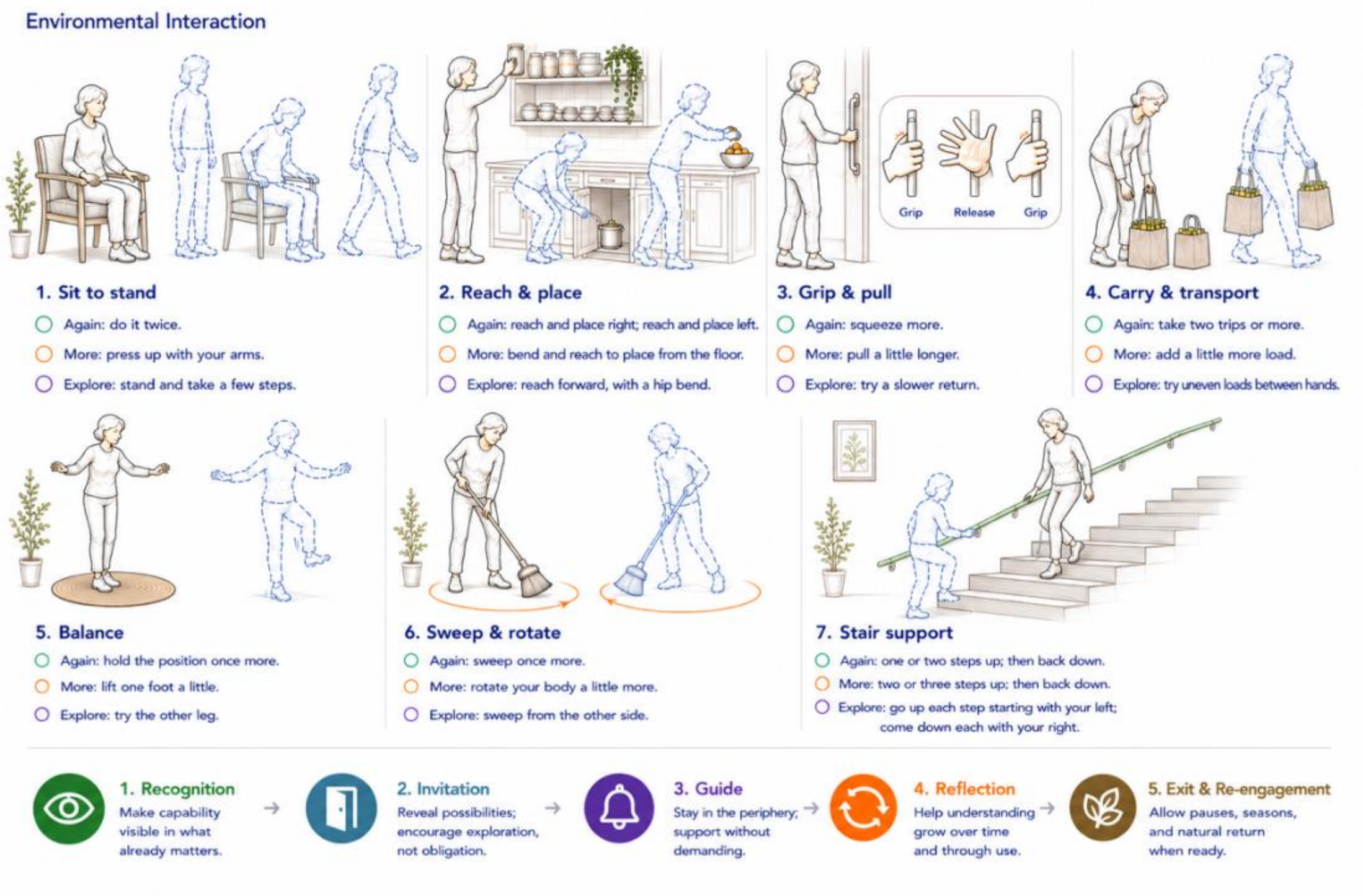

Figure 11: exploring further environmental interaction opportunities for the strength context.

### 5.1 Beneath invitation: the hidden work of recognition

Probe 1 confirmed that the system worked: participants followed the guides, performed the movements, and found the prompts intelligible. What it did not reveal was that recognition had already happened before any interaction with the system began. The onboarding conversation, the researcher presence, the in-lab setting had collectively established that everyday actions were worth attending to as strength practice. **The system inherited that foundation without creating it**.

The assumption was not unreasonable in context. Participants were self-selecting individuals who had expressed a desire to maintain strength and independence, and co-design sessions had surfaced positive responses to the reframing of everyday action as strength practice. What the probes revealed is that understanding a reframing in conversation and sustaining it as an ongoing orientation within everyday activity are different challenges — and that the accumulated weight of existing frameworks for what movement means, and what technology is expected to do, reasserts itself even when people have explicitly said they want something different. This is precisely what longitudinal in-situ deployment reveals that shorter or lab-based studies cannot.

Probe 2 made the cost of that assumption visible. Two of three participants reorganized the interaction as a set of tasks to complete, bunching their repetitions into the early morning rather than distributing them across ongoing activity. From a compliance perspective this looked like success — goals were met. But it was the opposite of the intended model. The guidance had been assimilated into an existing mental frame: technology gives tasks, tasks get done. Nothing in the system disrupted that frame, because the system assumed it had already been disrupted. This pattern resonates with prior work in HCI showing that users interpret system prompts within existing mental models of tasks and obligations [40, 60, 79] and with design approaches that emphasize openness and interpretation over control and optimization [51, 55, 85].

What the probes consistently suggest, across four deployments and a range of participants, is that this frame runs deeper than interaction design alone. Participants arrived with a settled understanding that strength is something you either have or have progressively lost — not something continuously expressed in the actions of daily life. This pattern is not unique to this study: research on self-perceptions of ageing suggests that negative beliefs about age and capability are not merely descriptive but actively shape behavior and physical outcomes [53, 103]. "I'm too old for X" is one surface form of this. The surprise that recurred across Probe 4 — "I hadn't thought of it that way," "that's it? just standing up?" — reflects the same underlying belief, encountered fresh. Notably, simply reframing existing activity as meeting a health standard has itself been shown to produce measurable change, without any modification of behavior [18] suggesting that recognition is not merely a precondition for engagement but may carry its own value. The reframing that everyday action already constitutes strength practice was genuinely news, each time it was introduced, to people who had been living those actions for decades. **Recognition, in this sense, is not a discrete event**, a single moment of reframing that, once achieved, persists. It is better understood as an ongoing orientation: a condition in which everyday action remains available as something more than functional transit. Sustaining that orientation is the design challenge.

This points to a design challenge the probes clearly named but did not resolve: how does the interaction itself establish recognition, rather than relying on external conversation to do so? How does a system communicate, in the moment of action, without a researcher present, that what you are already doing counts, and that you might want to explore it further? Designing for invitation presupposes that the person understands what they are being invited into. A prompt to extend action only makes sense once the action itself is understood as meaningful. Getting that foundation in place is the prior problem, and it is one that future work in this space will need to address directly.

### 5.2 Recognition before Extension

The probe sequence reveals a sequencing problem that is easy to miss. Extension — doing it twice, going further, exploring more — presupposes that the first instance of an action already means something to the person performing it. Where that meaning is absent, extension has no foundation. The action remains a task to complete rather than a capability to explore.

This sequencing was not built into the system. The "do it twice" prompt arrived without first establishing that the first time counted. In the co-design work and in Probe 4 sessions, that establishment happened through conversation and visibly mattered. In home deployments, it had to be assumed. The variation in uptake across Probes 2 and 3 is at least partly explained by this: participants for whom the reframing had landed, who had genuinely registered their everyday actions as strength work, engaged differently from those for whom it remained abstract.

The contribution is not that everyday action can be reframed: that can happen through conversation or demonstration, as this work shows. What the probes reveal is what follows: for some participants, a moment of recognition—'that's a strength move? I had no idea!'—carried forward into sustained engagement. For others, no such moment arrived, and without it the interaction defaulted to task completion. Recognition, when it came, opened the action rather than completed it — generating curiosity, playfulness, interest in what else might be possible. Designing for those conditions, rather than assuming recognition will arrive or persist, is the design challenge this work names. That is not an educational problem or a behavior change problem. It is an interactional one.

The design implication is sequential rather than stylistic. It is not enough to frame extension as invitation rather than instruction, though that matters too. The prior question is whether the system has first done the work of recognition, of making the action itself legible as something meaningful and worth attending to. Recognition establishes the ground. Extension is what becomes possible on it.

How a system might actively support that recognition, in the moment of action, without researcher or onboarding conversation to scaffold it, remains an open design question. But the probes make clear that it cannot be skipped, assumed, or delegated entirely to initial setup. It is part of the interaction.

### 5.3 From interaction as event to interaction as presence

Screen-based guidance, as used in earlier probes, proved effective for delivering structured prompts, but it also framed interaction as an event: something that appears, demands attention, and disappears. Probe 4 showed that this episodic quality was not simply a feature of screens but a representational stance that participants consistently pushed against. What they preferred was guidance that could exist as a persistent condition of the environment, available to be attended to or ignored, without requiring explicit activation or signaling its own presence. While HCI has extensively explored interaction through discrete events such as prompts, notifications [64], interruptions [45, 62], and task-oriented exchanges [41], less attention has been paid to interaction as a persistent condition of the environment rather than a sequence of individual events.

Ambient interfaces and calm technologies make system state or sensed information available at the periphery of attention [4, 61, 99, 101]. Environmental interaction instead makes the significance and possibilities of an action already underway continuously available for engagement. What persists is not information about activity, but an orientation toward activity in the moment. This is the sense in which environmental interaction differs from ambient systems, not in whether something persists, but in what persists.

This perspective builds on traditions in ubiquitous and embodied computing that embed sensing and response within everyday contexts [1, 98] and emphasize the situated, bodily nature of interaction [23]. Related work in data physicalizing has explored how representations of activity can persist in the environment to support reflection over time [47]. Environmental interaction differs in that the persistent element is not a representation of past action or system state, but the ongoing meaning of action as it unfolds.

Probe 4 also showed that presence is not simply a matter of persistence, it is a matter of representational fit. Figures, lights, sound, and scale all shaped whether guidance was experienced as available, intrusive, stigmatizing, or engaging. This includes not only functional clarity but how representations signal identity, capability, and social meaning. A form that is persistently present in the wrong register, as too clinical, too large, too explicit about its purpose, may be more disruptive than an episodic screen prompt. Designing for presence, as HCI and related literature has shown, requires attention to how a representation exists within the social and material texture of the home, not only whether it persists [17, 82, 95].

### 5.4 Separating doing and reflecting

A consistent distinction emerged from Probe 4 between supporting action in the moment and supporting reflection on action over time. Participants described these as distinct functions — and, importantly, this distinction was not introduced by the research team but arose from participants themselves as they engaged with different representational forms.

Attempts to combine both functions within a single interface risked obscuring each. Guidance for action works best when it is low-friction, situated, and ignorable — present in the environment, aligned with ongoing activity, easy to attend to or pass by. Reflection on action belongs in a different register entirely: explicit, inspectable, and intentionally entered. An app, accessed deliberately, serves that function in a way that an ambient environmental presence cannot and should not try to.

This extends prior work in personal informatics, which separates capture, reflection, and action into distinct phases [54]. What the probes add is a spatial and material dimension: these phases may also benefit from distinct physical and interactional registers, not merely distinct interface states within a single system. The doing interface and the reflecting interface are not just different modes — they may need to be different objects, in different places, encountered in different ways.

### 5.5 Expanding the option space over time

The probes raise a question that most HCI systems do not need to ask: what happens when interaction succeeds? If the goal is to help people recognize and extend the strength already present in their everyday actions, then success is not continued dependence on the system — but neither is it the system's withdrawal. The probe findings suggest a different frame entirely.

Participants' relationships with the system were not static. Some engaged actively early and settled into lighter use; others found the system more meaningful at certain times than others; and several — including participants in their eighties and nineties — noted that they would want reminders to remain available precisely because they knew their memory was not reliable. This is not a failure of competence. It reflects a sophisticated understanding of what the system offers: not a scaffold to be removed once capability is established, but an expansion of the option space — a persistent availability that supports choice rather than dependence.

The cane is a useful analogy here. The goal of strength-building is not to make the cane unnecessary; it is to ensure that on days when you want to walk without it, you can, and on days when you want it, it is there. Environmental interaction systems may work similarly: as recognition and practice develop, the nature of engagement changes — from active guidance toward ambient availability — but the system remains part of the environment rather than being withdrawn from it. , and this differs from scaffolding models common in HCI, which tend to assume that successful systems are eventually outgrown [76, 102], and from engagement models that emphasize sustained interaction with a system over time [73].

This evolution is also shaped by context in ways the probes made consistently visible. Engagement varied not only with system design but with the organization of everyday routines, prior understandings of movement and activity, physical constraints of the home, and social circumstances such as cohabitation. The same system could be taken up as an embedded part of ongoing activity by one participant and reorganized as a discrete morning task by another — not because the system behaved differently, but because the contexts into which it arrived were different. Environmental interaction is in this sense constituted through context: it does not simply operate within everyday life but is shaped by it, in ways that cannot be fully anticipated in advance and that vary across people, households, and moments in time.

Designing for this requires attention not only to interface and representation but to how interaction will be encountered against the grain of existing routines, beliefs, and living arrangements — and how it can remain available and meaningful across the full range of those encounters, rather than assuming a single trajectory of use.

### 5.6 Toward environmental interaction

Taken together, the probes trace a consistent arc. What began as a question about how to support movement without adding new tasks revealed a deeper set of challenges: that recognition of everyday action as meaningful cannot be assumed; that guidance lands differently depending on the mental models and beliefs people bring to it; that presence and fit matter as much as functionality; and that the relationship between a person and a system evolves over time in ways that neither scaffolding nor sustained-engagement models fully capture.

These findings locate a space that sits between existing HCI approaches rather than extending any one of them. Movement support systems add activity; monitoring systems observe it; tangible and embodied systems center engagement with designed artefacts. Environmental interaction shifts focus: it makes legible what is already happening, and opens the possibility of engaging with it differently. The design challenge is not to introduce interaction into the environment but to reveal the interaction already present there.

Across the probe sequence, something consistent emerged about the conditions surrounding recognition. In Probe 1, participants continued interacting with the CanBand beyond any prompted sequence, drawn by curiosity, repeating actions to see what would change. In Probe 4, participants responded to light- and sound-augmented forms with delight, already imagining how they would personalize them, not because anything was wrong, but because the interaction felt worth making their own. In both cases, the engagement was exploratory rather than compliant. People were drawn toward the interaction rather than directed through it. Whether this quality of engagement is sufficient to support recognition on its own, or whether the reframing of everyday action as meaningful capability requires something additional, the probes do not resolve. What they do suggest is that playfulness and curiosity-driven engagement create the conditions in which recognition might arrive; that delight may open the moment rather than fill it. Designing for that opening, for an interaction that makes a person curious enough to discover what they are already doing, is perhaps the most generative direction this work reveals.

What the probe sequence also showed, and what none of the in-lab or short-term studies that preceded it could have shown, is that this revelation is not a one-time event. The surprise of discovering that a chair sit-stand is a box squat, that pulling open a door is a one-arm row, recurred across every context in which it was introduced. This suggests that *the reframing itself is part of the interaction, not just its precondition*. Systems in this space may need to keep making that legibility available, not (just) as onboarding, but as a persistent quality of the environment.

The design space extends beyond strength-building. The principle is not domain-specific: wherever everyday life contains actions whose capability-bearing potential goes unrecognized, from how we organize social attention, prepare food, or move through the quieter transitions of the day, the same design challenge applies. Not to add new practices, but to make legible what is already there, and to do so in ways that expand what people understand themselves to be capable of, rather than simply adding to what they are asked to do. Everyday environments are already full of interactions waiting to be recognized. The design challenge is not to introduce interaction into the environment, but to make visible what is already, in the inter-action, present within it.

## 6 LIMITATIONS

This work is exploratory in scope, using probe-based deployments to investigate environmental interaction as an emerging design space within HCI. It is not intended to establish generalizable behavioral or physiological outcomes, but to characterize how environmental interaction is interpreted, taken up, and shaped within everyday contexts.

A further limitation concerns the locus of recognition in these deployments. Across the probes, the reframing of everyday action as meaningful capability arrived through researcher conversation and demonstration rather than through the system itself. The system supported detection, guidance, and extension; it did not support recognition as an ongoing condition of the interaction. This is both a limitation and, we would argue, the most generative finding: *it names a design challenge the probes could not yet address: how interaction might sustain that orientation toward action as openly available, rather than assuming it or delegating it to onboarding*. That question is precisely what motivates the future work directions identified in Section 7.

The work presented here centers on supporting strength to sustain independence in older adults. This focus therefore limits what can be said about how environmental interaction may be instantiated in other populations or settings. Observations during the probes suggest that this design space is be extensible beyond this context; however, as such applicability is not examined here, this remains an open invitation for future work.

The probes reported here explore a subset of everyday actions: primarily sit–stand movement and related interactions. These do not capture the full range of possible environmental interactions, including ones for which we have developed prototypes including balance, grip, pulling and pushing (as per Figure 1). Findings should be understood in relation to this focus. Variation across participants further shows that interaction in this space is contingent on individual circumstances, routines, and interpretations, rather than uniformly taken up across contexts.

The methodological approach also sets limits on what can be assessed. While probe-based deployments enable rich insights into lived experience, they do not provide the level of scale or longitudinal evidence required to assess effectiveness in terms of health outcomes over time, which remain important directions for future investigation.

## 7 FUTURE WORK

This work opens several directions for further research into environmental interaction as a design space. A key opportunity is for future work to investigate how recognition is established, maintained, and sustained across different domains of everyday action.

**Representation and presence** A first direction concerns the role of physical and ambient interfaces as persistent elements within the environment. Findings from Probe 4 suggest that interaction may be more effectively supported through forms that are continuously present rather than episodically delivered through screens. Future work should explore how movement guidance can be represented in ways that are both perceptible and unobtrusive, and how such representations balance persistence and immediacy—for example, through static presence, dynamic sequences, or context-sensitive responses that unfold at the moment of action.

**Trajectory over time** A second direction concerns how interaction evolves over time, including how environmental interaction supports building knowledge, skills and practice. Rather than assuming that support is gradually withdrawn (e.g. scaffolding and fading), our findings suggest examining how interaction contributes to the translation of movement practices across contexts, such as from the home into everyday activity beyond it. This raises questions about whether representations remain stable, transform, or cycle overtime, and what constitutes an evolution of environmental interaction over time. This also raises questions about how interaction remains available without becoming obligatory, to support trajectories in which engagement can be taken up, set aside, and re-engaged over time, rather than withdrawn entirely.

**Environment vs wearable systems** A related direction concerns the relationship between environmental and wearable approaches to supporting movement. This work foregrounds the environment as the primary site of interaction, reducing reliance on personal devices. Further research is needed to understand when environmental approaches are sufficient, how they complement wearable systems, and what forms of interaction are enabled or constrained by each.

**Reflection and lived experience** A further direction concerns the role of reflection and lived experience. Participants highlighted the importance of reflecting on activity in relation to how it is experienced and felt, rather than solely through abstract tracking. Future work could explore how environmental interaction supports awareness of effort, capability, and change over time, building on work such as inbodied interaction [86] and tuning [87].

**Designing for the recognition moment** This work reveals recognition of everyday action as meaningful capability as a distinct and prior design challenge. But the probes also show something more: when that moment of discovery arrives, *it is experienced as delight rather than information*. It is a moment that opens rather than instructs. Future work opportunities include exploring how interaction might actively support and sustain this moment; how environments might be designed not to prompt action, which can fall into the Additional Task interpretation, but to make that moment of inter-action and its capability-bearing potential discoverable, playful, and worth returning to. This is a different design challenge from guidance or tracking, and it may be the most generative direction this work opens.

**Extending the design space** Finally, this work focuses on a limited set of actions and contexts. Extending environmental interaction to a broader range of activities, environments, and populations remains an important direction, including how different forms of movement, capability, and domestic or shared environments shape how interaction is understood and taken up.

Interest in the system extended beyond the recruited participant group, including, notably in the Probe 4 group, from an individual using a wheelchair for rehabilitation, wanting to have the system at her workplace. This suggests that the principle of recognizing and extending actions already underway can have relevance across a broader range of environments than explored here. Taken together, these directions point toward environmental interaction not as a single system or technique, but as an evolving design space that raises new questions about how interaction is embedded in, and emerges from, everyday life. Where the site of the interaction *is* the interaction.

## 8 CONCLUSION

This paper began with a question about movement, but ended somewhere different. The probes were designed to explore whether everyday actions could be recognized and extended as strength practice within domestic environments. What they revealed was something less expected: the locus for interaction. The system could detect sit-stand transitions, offer guidance, and run reliably in people's homes. What it did not do — and what no system yet does well — is establish the foundational recognition that makes any of that matter. The discovery that a chair sit-stand is a box squat, that pulling open a door is a row, that "you are already doing this," landed as genuine news, every time, to people who had been performing those actions for as long as they had been doing these actions; and it landed not as instruction but as *delight*: a moment of recognition that opened rather than directed. That is the finding the probes produced that no lab study focused on our usual question of "does it work" could have reached.

Environmental interaction names the design space this opens. It is not a system or a technique but a reframing of where interaction is located, and of what it is for in this context. Where prompting systems say "do this now," and tracking systems say "here is what you did," environmental interaction says something different: its focus is "you are already in the action, and there are choices available in this moment." The role of the system is not to direct behavior toward a predetermined outcome. Instead, its intent is to make visible what is already happening, expanding what the person understands themselves to be capable of doing with it. Environmental interaction re-situates: what designers look for, what systems need to do, and how success is understood. The principle is not domain-specific: wherever everyday life contains actions whose capability-bearing potential goes unrecognized — in how we organize attention, move through social spaces, or approach the quieter transitions of the day — the same design challenge applies.

In environmental interaction, the site of interaction is not primarily a device, prompt, or monitor, but t*he action already underway between person and environment.* Sitting, standing, reaching, lifting, stabilizing, these are not treated merely as behaviors to count or tasks to trigger; they become moments that can be recognized, valued, and optionally extended in situ. Rather than measuring action from outside or adding a new task to complete, environmental interaction works within what is already taking place, oriented toward making capability legible in the moment. This in-the-action-in-the-moment-exploration locus is a design aspiration the probes both advanced and, in revealing this recognition challenge, more precisely defined.

The design challenge is therefore not to introduce interaction into the environment, *but to recognize and support the interaction already present there* — and in doing so, to expand what people understand themselves already to be doing. Environmental interaction offers one way of approaching this challenge, drawing attention to how actions already embedded within everyday environments can become sites of recognition and extension. The probes suggest that the challenge is not only how to support action, but how to support recognition of the capability already present within everyday action.

## ACKNOWLEDGEMENTS

UKRI EPSRC grants – Incidental Interaction: Novel Technology to Support Elders-as-Athletes through Augmenting Everyday Interactions and Health Resilience Interactive Technology: transforming self-management for individual and community health via inbodied interaction design

With thanks as well to all our participants, and our engagement as well with Hampshire Council Social Carers.